\documentclass[preprint,showpacs,preprintnumbers,amsmath,amssymb]{revtex4}

\usepackage{graphicx}
\usepackage{dcolumn}
\usepackage{bm}
\begin{document}

\title{Coulomb excitation of $^{11}$Be reexamined}

\author{T. Tarutina}
\email{tatiana@fma.if.usp.br}
\author{L.C. Chamon}
\author{M.S. Hussein}
\affiliation{Departamento de F\'isica Nuclear, Instituto de F\'isica
  da Universidade de S\~ao Paulo, \\ 
Caixa Postal 66318, 05315-970 S\~ao Paulo, SP, Brazil}

\date{ 28 November 2002}

\begin{abstract}
The study of Coulomb excitation of $^{11}$Be to the first excited
state in intermediate energy collisions with heavy targets is presented.
The existing experimental data are reanalised by including the
probability of projectile survival into the calculation of Coulomb
excitation cross sections. The survival probabilities are
calculated using a recently developed global optical model potential
tailored in line with the double folding model. 
The extracted
$B(E1)$ values for the transition $\frac{1}{2}^+\longrightarrow
\frac{1}{2}^-$ in $^{11}$Be are found to be slightly larger than
those obtained so far using the $b_0$-recipe in Coulomb excitation
theory. 
\end{abstract}

\pacs{27.20.+n, 25.60.-t, 25.70.De}

\maketitle

The Coulomb excitation process has  been proven to be a useful
tool to study nuclear structure (see, {\it e.g.}, Ref.\cite{Ber01}). 
It is an attractive test of the
different nuclear models as the Coulomb interaction is very well
known. Recently a number of experiments has been done to study the
structure of light neutron rich nuclei, such as
$^{11}$Be \cite{Ann95,Nak97,Fau97}, $^{15}$C, $^{17}$C, $^{19}$C
and several oxygen isotopes, using the Coulomb excitation process
\cite{Pra02}.

The isotope $^{11}$Be is one of the most studied halo nuclei
nowadays. The reason why $^{11}$Be is an attractive system to
study by Coulomb excitation lies in the peculiar structure of the
spectrum of this nucleus. Both  the $\frac{1}{2}^+$ ground state
and the only {\it bound} excited state $\frac{1}{2}^-$  are weakly bound
and have energy difference of 0.32 MeV, thus making the $E$1
transition in $^{11}$Be the fastest known between bound states.

There have been several experiments that measured  the excitation
of the first excited state in $^{11}$Be and extracted the
corresponding $B(E1)$ value. The adopted value {\it ca}
0.116 $e^2$fm$^2$ was obtained by Millener {\it et al.} in
Ref.\cite{Mil83} by averaging the results of three experiments on
the lifetime of the excited state using a Doppler-shift technique.

The experiment done in GANIL \cite{Ann95} studied the inelastic
scattering of $^{11}$Be on a lead target at a bombarding energy of
45 MeV/u. The obtained cross section was only 40$\%$ of the one
predicted by calculations for pure Coulomb excitation. To explain
this large difference it was proposed that the higher order
effects may contribute, but calculations done in Ref.\cite{Ber95}
showed that the cross section falls by only 4$\%$ if coupling to
continuum was taken into account. The inclusion of monopole and
quadrupole modes of nuclear excitation also performed in the work
\cite{Ber95} resulted in the slight increase of the cross section
less than 2$\%$.

Similar experiment was done by Nakamura {\it et al.} in
Ref.\cite{Nak97} at $E$ = 64 MeV/u. The $B(E1)$ value extracted
from the cross section assuming pure Coulomb excitation was
comparable with the result of the Millener analysis. Later the
inelastic scattering of $^{11}$Be was measured at MSU \cite{Fau97}
 on lead and gold targets for energies 60 and 58 MeV/u, respectively.
The extracted $B(E1)$ value confirmed that of \cite{Nak97} and agreed,
at least marginally, with the lifetime experiments. For summary of
this see Table \ref{table_results}.

In the analysis performed in the mentioned papers, the authors
used the formalism of pure Coulomb excitation and excluded nuclear
processes approximately by using a low impact parameter cutoff
$b_0$. In this work we show that the inclusion of the full fledged
survival probability $|S(b)|^2$ in conjunction with $b_0$, though
slightly improves the agreement with the data, it does allow one
to perform a more realistic smooth cutoff based analysis.

In the calculation of the Coulomb excitation cross section, the
important ingredient of the model is the minimum value of the
impact parameter, $b_0$,  from which the integration of the
excitation amplitude is performed. This $b_0$ is roughly
determined by the sum of the projectile and target radii.
$^{11}$Be, being a halo nucleus, has a very diffuse structure,
which makes the definition of the minimum impact parameter
obscure.

To improve upon the calculation we multiply  the probability of
the Coulomb excitation at a given impact parameter by the survival
probability  $|S(b)|^2$ calculated for the system under
consideration. In this work we take the S-matrix from the optical
model. The calculation of the elastic S-matrices requires the knowledge of
the densities of the projectile and the target. The halo density of
$^{11}$Be was calculated in the framework of the particle-rotor
model, which includes the excitation of the rotational 2$^+$ state
of the core $^{10}$Be \cite{Nun96}.

 As was mentioned above, the contribution of nuclear excitation is small.
This is so since nuclear effects are limited to a
very small impact parameter region around the grazing value. In fact,
as shown in Refs.\cite{Mot94,Ber94,Shy96} the Coulomb excitation was 
found to be by far
the dominant piece of the cross section. Further, in Ref.\cite{Fau97}
an experiment was done with light targets
to study the importance of nuclear excitation. The results were
4.0 mb for carbon target and 1.7 mb for beryllium target. Since the
nuclear cross section corresponds to the area of a ring around grazing 
$b_0$, it scales with $A$ as $A^{1/3}$. Accordingly, we find for the
lead target approximately 10 mb nuclear excitation contribution implying a
mere 3$\%$ effect. Therefore in the following we ignore the nuclear
excitation effect.

The semiclassical model is usually used to describe Coulomb
excitation at intermediate and high energies. This model assumes
the straight line trajectory for the projectile and treats quantum
mechanically  the absorption of radiation by the nucleus. The
formalism of this method was presented in Ref.\cite{Ald75} and
later was extended to the relativistic case in Ref.\cite{Win79}.
For the high energies the first order perturbation theory is a good
approximation to calculate the amplitudes for Coulomb excitation.
In the first order perturbation theory the process of Coulomb
excitation can be described as emission and absorption of virtual
photons \cite{Fer24}.
We included
only $E1$ multipolarity in our analysis since for the 
transition $\frac{1}{2}^+\longrightarrow\frac{1}{2}^-$ in $^{11}$Be 
the dipole multipolarity is the dominant one \cite{Ber88}. 

Using the formalism of virtual photons the Coulomb excitation
cross section has the following form:

\begin{equation}
\sigma_C= \int_0^{\theta_0}\frac{d\sigma_{C}}{d\Omega}d\Omega=
2\pi\int_{b_0}^{\infty}P(b)bdb \label{cross}
\end{equation}
with
\begin{equation}
P(b)=\frac{16\pi^3}{9\hbar c}B(E1)N_{E1}(\omega,b), \label{prob}
\end{equation}
and $b_0=a\cot{\frac{\theta_0}{2}}$,
where $a$ is half the distance of closest approach for head-on
Coulomb collision. $N_{E1}(\omega,b)$ is the number of virtual 
photons given by
\begin{equation}
N_{E1}(\omega,b)=\frac{Z_1^2\alpha}{\pi^2}\left(\frac{\xi}{b}\right)^2\left(\frac{c}{v}\right)^2\left[K_1^2\left(\xi\right)+ \frac{1}{\gamma^2}K_0^2\left(\xi\right)\right], \label{phot}
\end{equation}
where $Z_1e$ is the charge of the target, $\xi = \frac{\omega
  b}{\gamma v}$, $\omega=E_{ex}/\hbar$,
$E_{ex}$ is the excitation energy of the state in the projectile,
$v$ is the relative energy, $\gamma$ is the relativistic factor
and $\alpha$ is the fine structure constant. The behavior of the
integrand in the Eq.(\ref{cross}) is determined by the impact
parameter dependence of the modified Bessel functions of the zero
and first order $K_0$ and $K_1$.

Coulomb recoil was taken into account using the method of Winther
and Alder \cite{Win79} replacing  $b$ in the expression for
$P(b)$ by $b'=b+\pi a/2\gamma$. 
Further, we multiply $P(b)$ of
Eq.(\ref{cross}) by the survival probability $|S(b)|^2$,
calculated from the optical potential NLM3Y of Ref.\cite{Cha02}
and then write
\begin{equation}
\tilde P(b)=P(b)\cdot|S(b)|^2.
\end{equation}

The optical potential was calculated from the double folding
model, using projectile and target densities and the effective
nucleon-nucleon interaction. The effect of the exchange-related
non-locality is taken fully into account. The local-equivalent,
energy-dependent potential can be written
as \cite{Cha02}
$$
V(r,E)=V_N(r,E)-iW(r,E),
$$
where
$$
V_N(r,E)=V_F(r)e^{-4v^2/c^2},
$$
where $c$ is the speed of light and $v$ is the local relative
velocity between the two nuclei
\begin{equation}
v^2(r,E)=\frac{2}{\mu}(E-V_C(r)-V_N(r,E)).
\end{equation}

The local folding potential $V_F(r)$ is obtained in the usual way.
For the Coulomb part, $V_C(r)$ we employ a similar folding
prescription using the proton densities. The imaginary part of the
optical potential was found to be
\begin{equation}
W(r,E)\approx 0.78V_N(r,E).
\end{equation}
The determination of the number 0.78 is discussed in
Ref.\cite{Alv02}. This number was obtained by fitting 
calculated elastic scattering
cross sections with experimental data for the following systems:
$^{12}$C+$^{12}$C,$^{16}$O,$^{40}$Ca,$^{90}$Zr,$^{208}$Pb; 
$^{16}$O+$^{208}$Pb, $^{40}$Ar+$^{208}$Pb
and for the wide range of energies. The calculations showed that the
number 0.78 is approximately system-independent.

We employ the same potential here by using the appropriate
halo+core density of $^{11}$Be. For the densities of the heavy targets
and the two-parameter Fermi distributions were used
\begin{equation}
\rho(r)=\frac{\rho_0}{1+\exp{\frac{r-R_0}{a}}},\label{glob_dens}
\end{equation}
where $a=0.56$ fm, $R_0=1.31A^{1/3}-0.84$ fm and $\rho_0$ is determined 
by the normalization condition.  These densities were obtained in 
Ref.\cite{Cha02} with the aim of providing a global description of the
nuclear interaction, based on an extensive study involving charged
distributions extracted from electron scattering experiments and 
theoretical densities calculated through the Dirac-Hartree-Bogoliubov model.
For the density of the core of $^{11}$Be, $^{10}$Be, we used the
Eq.(\ref{glob_dens}) with $a=0.5$ fm and $R_0=1.75$ fm, which gives
$R_{rms}$($^{10}$Be) = 2.3 fm.

We use the existing particle + core excitation model to describe
the halo structure of a light halo nucleus such as $^{11}$Be
\cite{Nun96}. The inclusion of the core excitation allows for the
coupling between the collective degrees of freedom of the core and
the orbital motion of the single neutron. In this calculation we
assumed a quadrupole-deformed core (with deformation parameter
$\beta_2$) and
included only the ground state 0$^+$ and the first excited
rotational state 2$^+$.
The interaction between the valence particle and the core was
described by a deformed Woods-Saxon potential. The spin-orbit
interaction, $V_{so}$, is the standard undeformed one.
To reproduce the adopted $B(E$1) of $^{11}$Be we adjusted the
model parameters obtained in Ref.\cite{Nun96} and these parameters
are presented in Table \ref{table_pc}. The particle-core model with these
parameters gives a radius for $^{11}$Be of 3.0 fm, the
$B(E1)=0.116$ $e^2$fm$^2$.

\begin{table}
\centering
\begin{tabular}{ccccccc}\hline\hline

 $R_0$ & $a$ & $V_{so}$ & V$_{ws}^{EVEN}$ &
 V$_{ws}^{ODD}$ & $\beta_2$ & $E^*$(2$^+$)\\
 fm & fm & MeV & MeV & MeV &  & MeV \\

\hline
2.483 & 0.65 & 5.0 & 54.65 & 47.82 & 0.67 & 3.368 \\
\hline
\hline
\end{tabular}
\caption{\label{table_pc}Parameters of the particle-core model used to
calculate density of $^{11}$Be. $V_{so}$ and $V_{ws}^{EVEN/ODD}$ are 
the strengths of spin-orbit and Woods-Saxon potentials, $a$ and $R_o$
are the diffuseness and the radius of the potentials. $\beta_2$ is the
quadrupole deformation parameter and $E^*(2^+)$ is the excitation
energy of the core.}
\end{table}

\begin{figure}
\centerline{\includegraphics[scale=0.5]{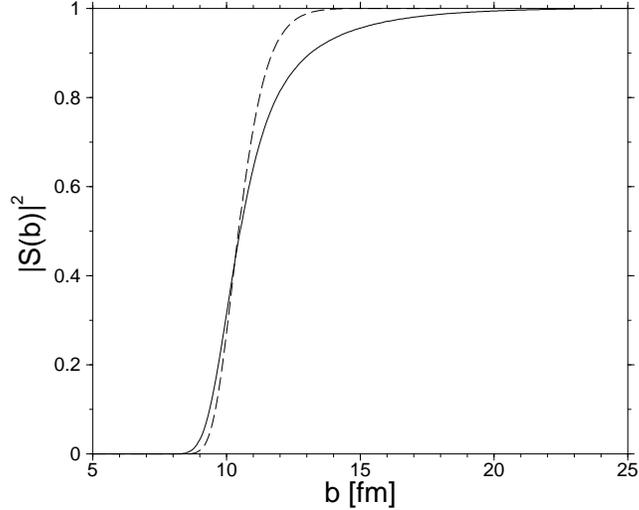}}
\caption{The projectile survival probability $|S(b)|^2$ calculated
for $^{11}$Be + $^{208}$Pb at 59.7 MeV/u. The solid line shows
the projectile survival probability calculated using the densities 
of the halo, dashed line stands for the calculation using the 
global densities of the Eq.(\ref{glob_dens}).}
\label{fig_smat}
\end{figure}

In Fig.\ref{fig_smat} we show the calculated squared moduli of
the elastic S-matrices for $^{11}$Be + $^{208}$Pb at 59.7 MeV/u. 
It is seen that using the density with the halo reduces 
the survival probability.

\begin{figure}
\centerline{\includegraphics[scale=0.5]{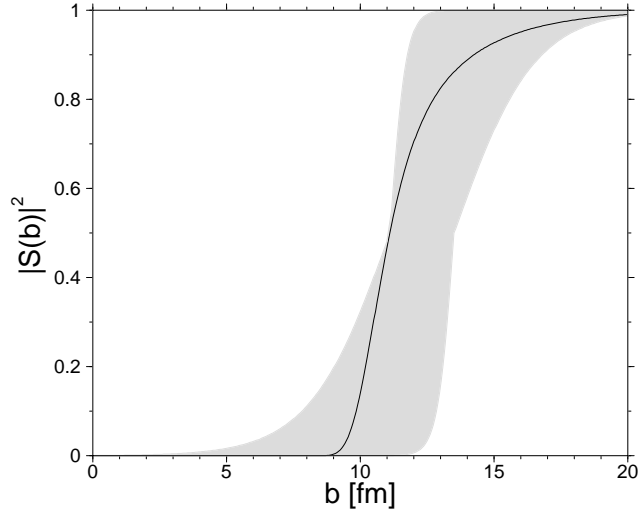}}
\caption{The projectile survival probability $|S(b)|^2$ calculated
for $^{11}$Be + $^{208}$Pb at 72 MeV/u including halo density. 
The shaded area represents 
the projectile survival probability determined from experiment 
in \cite{Nak94}.}
\label{fig_nak}
\end{figure}

In Ref.\cite{Nak94} the projectile survival probability was determined from
the angle dependence of the Coulomb dissociation cross section. The very
small number of experimental points (3 in the grazing region) with
their rather large error bars is certainly insufficient to make
meaningful comparison.
In Ref.\cite{Nak94} the survival probability was approximated by the 
function with the shape $1/(1+\exp{(-(b-b_0)/a)})$ with
parameters $b_0=12.3(1.2)$ fm and $a=0.9(0.6)$ fm. In fact, one
should include the size of the error bars in the presentation of the
experimental survival probability. For the purpose of completeness
we show in Fig.\ref{fig_nak}  the experimentally determined survival
probability of Nakamura {\it et al.} presented in the shaded area
together with ours shown as the full line. It is clear that our
calculation of the survival probability agrees reasonably well with 
Nakamura's one.

The experimental data on Coulomb excitation of $^{11}$Be is
summarized in Table \ref{table_results}. The fifth column shows
the values of $B(E1)$ extracted from the corresponding Coulomb
excitation cross section by assuming pure Coulomb excitation
and choosing appropriate $b_0$, that
is without including projectile survival probability. In the analysis
of Nakamura {et al.}\cite{Nak97} Eq.(\ref{cross}) was used
with $b_0=12.3$ fm which was determined from the
impact parameter dependence observed in the Coulomb breakup of
$^{11}$Be \cite{Nak94}. 
Fauerbach {\it et al.} in Ref.\cite{Fau97}
used the similar formalism to extract $B(E1)$ values from
the experimental cross sections with $b_0$ obtained from the aperture
of the experimental setup.

In our calculation we use $\tilde P(b)$ in Eq.(\ref{cross}) and
start the integration using the experimental value of $b_0$, which
is determined by the aperture of experimental setup. Thus for the
case of, {\it e.g.}, $^{11}$Be + $^{208}$Pb at 59.7 MeV/u we
integrated the probability shown by dotted line starting with
$b_0=11.5$ fm.

\begin{figure}[t]
\centerline{\includegraphics[scale=0.5]{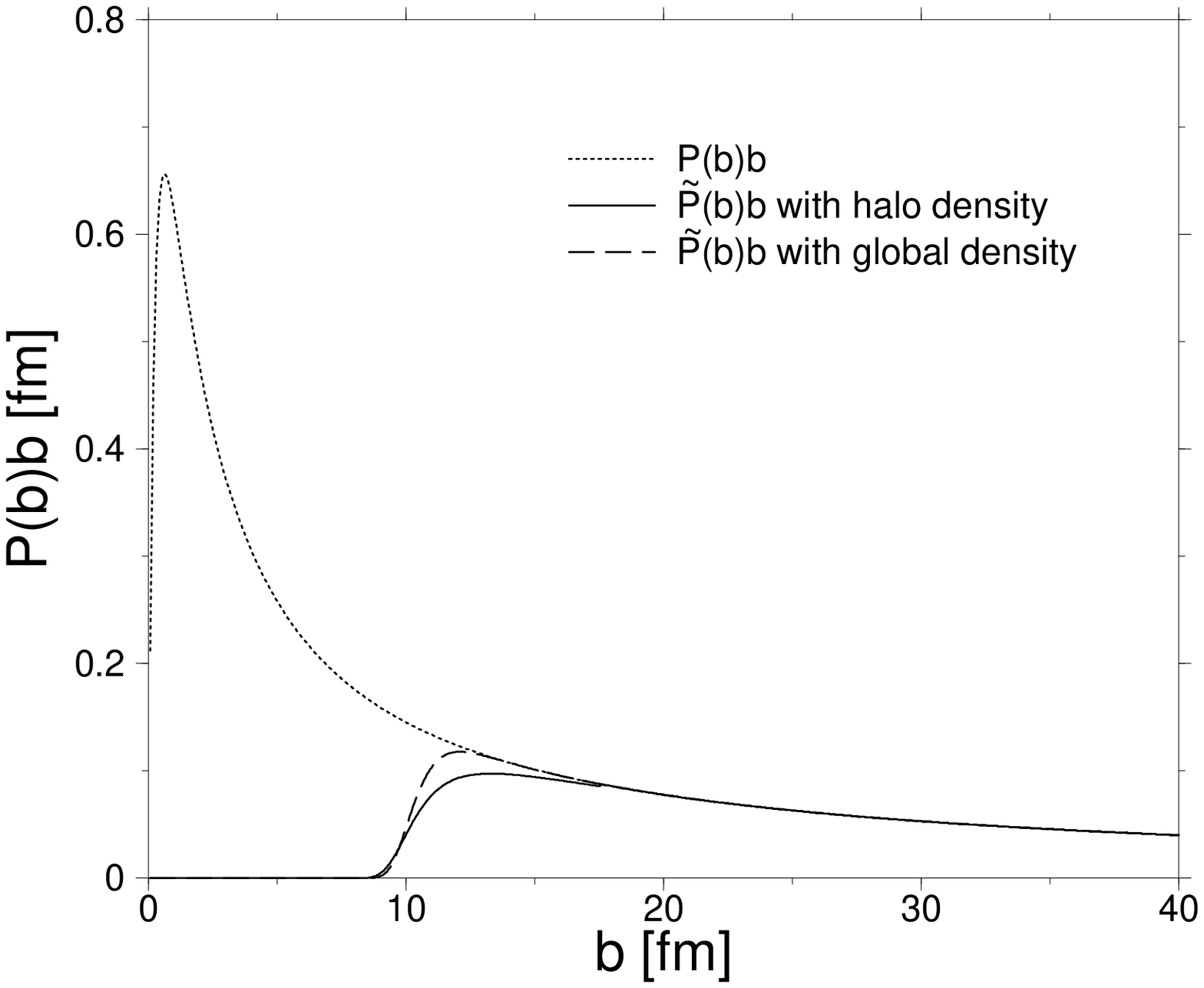}}
\caption{The impact parameter dependence of $P(b)b$ and $\tilde
 P(b)b$. The case of $^{11}$Be + $^{208}$Pb at 59.7 MeV/u is
 considered. $\tilde P(b)b$ was calculated with the halo density
 (solid line) and with global densities of the work \cite{Cha02}
 (dashed line) in $|S(b)|^2$, see text for details.}
\label{fig_int}
\end{figure}

\begin{table*}[t]
\centering
\begin{tabular}{ccccccccc}\hline\hline
Ref. & Target & Beam energy & $\sigma_C$ (exp.) & B($E$1) (old)&
     $\sigma_C$ (th. (1)) & $\sigma_C$ (th. (2)) & $\sigma_C$ (th. (3)) &
     B($E$1) (this work)\\
     &        & MeV/u &  mb      &  $e^2$fm$^2$ & mb & mb & mb & $e^2$fm$^2$
\\
\hline
\cite{Ann95} & $^{208}$Pb & 45 & 191(26) & 0.047(06) & 197  & 193 &
     188 &0.048(06)\\
\cite{Nak97} & $^{208}$Pb & 64 & 302(31) & 0.099(10) & 314 & 309 & 
             303 & 0.099(10)\\
\cite{Fau97} & $^{208}$Pb & 59.7 & 304(34) & 0.094(11)& 304 & 295 &
             291 & 0.098(11)\\
\cite{Fau97} & $^{197}$Au & 59.7 & 244(25) & 0.079(08) & 244 & 236 &
     233 &0.083(08)\\
\hline
\hline
\end{tabular}
\caption{\label{table_results} The experimental and theoretical 
Coulomb excitation cross sections for different energies and extracted $B(E1)$.
The fifth column shows the $B(E1)$ extracted from experimental $\sigma$
using pure Coulomb excitation in Refs.\cite{Nak97,Fau97}. Columns 6,7
and 8 show calculated
cross section (using $B(E1)$ from column 5) assuming no Coulomb recoil
and no halo-type density
(th. (1)); with Coulomb recoil, but no halo-type density (th. (2));
with Coulomb recoil and halo-type density (th. (3)).
The last column shows the revised values of $B(E1)$ which
reproduce experimental cross section using $\tilde P(b)$ and taking
Coulomb recoil into account.}
\end{table*}

The theoretical cross sections calculated in this work are presented 
in columns 6, 7 and 8 of Table \ref{table_results}.  
First we assumed the $B(E1)$ values from the previous analysis 
\cite{Nak97,Fau97} and
calculated Coulomb excitation cross sections without taking into
account Coulomb recoil and using density of Eq.(\ref{glob_dens}).
We obtained the cross sections slightly larger than the experimental
for $^{11}$Be + $^{208}$Pb at 64 and 45 MeV/u as we integrated
the probability from $b_0$ obtained from experimental aperture
and not from grazing impact parameter as was done in \cite{Nak97}.
Taking into account Coulomb recoil reduced the calculated cross
section by approximately 2$\%$. The further reduction (approximately
2$\%$) in the calculated cross section was obtained by using the
halo densities of $^{11}$Be.

In Fig.\ref{fig_int} we show the impact parameter dependence of
the Coulomb excitation probabilities $P(b)$ and $\tilde P(b)$ 
for the case of $^{11}$Be + $^{208}$Pb at 59.4 MeV/u. The
inclusion of projectile survival probability reduces the Coulomb
excitation probability for impact parameters from 11.5 fm and
up to 17 fm.
This results in the reduction of the cross section by 4$\%$ compared 
to the result of the pure Coulomb
excitation with $b_0=11.5$ fm. For the case of $^{11}$Be + $^{197}$Au  the 
the reduction of the calculated cross section was 5$\%$.

Finally, we extracted the revised values 
of $B(E1)$ using $\tilde P(b)$ and taking Coulomb recoil into
account. We found no difference between the values of $B(E1)$
extracted in the work \cite{Nak97} and results of our analysis.  The
revised $B(E1)$ values obtained from the experimental cross sections
of the work \cite{Fau97} are approximately 4$\%$ larger than those
obtained in \cite{Fau97}. The revised $B(E1)$ values are presented in
the last column of the Table \ref{table_results}.

In conclusion, a better treatment of nuclear
absorption exemplified by the use of an appropriate survival
probability
in the calculation of the Coulomb excitation cross
section of $^{11}$Be leads to slightly smaller cross-section and then
a slightly (about 4$\%$) larger $B(E1)$ value. We expect similar 
effect in the Coulomb dissociation cross-section of $^{11}$Be and
other halo nuclei.

This work was supported in part by FAPESP and the CNPq.

\end{document}